# Variations of the Respiration Signals for Respiratory-Gated Radiotherapy Using the Video Coached Respiration Guiding System

Hyun Jeong Lee

*Department of Physics, Yeungnam University, Gyeongsan*

Ji Woon Yea

*Department of Radiation Oncology, Yeungnam University College of Medicine, Daegu*

Se An Oh

*Department of Radiation Oncology, Yeungnam University Medical Center, Daegu*

Respiratory-gated radiation therapy (RGRT) has been used to minimize the dose to normal tissue in lung-cancer radiotherapy. The present research aims to improve the regularity of respiration in RGRT using a video coached respiration guiding system. In the study, 16 patients with lung cancer were evaluated. The respiration signals of the patients were measured by a real-time position management (RPM) Respiratory Gating System (Varian, USA) and the patients were trained using the video coached respiration guiding system. The patients performed free breathing and guided breathing, and the respiratory cycles were acquired for ~5 min. Then, Microsoft Excel 2010 software was used to calculate the mean and standard deviation for each phase. The standard deviation was computed in order to analyze the improvement in the respiratory regularity with respect to the period and displacement. The standard deviation of the guided breathing decreased to 65.14% in the inhale peak and 71.04% in the exhale peak compared with the values for the free breathing of patient 6. It was found that the standard deviation of the respiratory cycle was decreased using the respiratory guiding system.

The respiratory regularity was significantly improved using the video coached respiration guiding system. Therefore, the system is useful for improving the accuracy and efficiency of RGRT.

Email: sean.oh5235@gmail.com

Fax: +82-53-624-3599

# I. INTRODUCTION

The purpose of radiation therapy is to supply as much radiation as possible to the target volume while minimizing the dose to the normal tissue [1]. The development of radiation therapy recently saw the emergence. The typical kinds of therapy are intensity-modulated radiation therapy (IMRT) [2–5] and image-guided radiation therapy (IGRT) [6–7]. Various techniques consider the movement of the internal organs in respiratory radiation therapy: (1) respiratory-gated radiation therapy (RGRT) irradiates only the tumor in a limited range [8–13], (2) tumor-tracking radiation therapy irradiates the tumor by tracking its movement [14–15], (3) breath-hold radiation therapy controls the respiration of the patient by intent [16–17], (4) abdominal compression uses a pressure device on the patient's abdomen [18–19], etc.

Receive the significant effect on respiration for tumor in the chest and the abdomen regions at least 2 cm, depending on the patient's position and the tumor's displacement [8, 20–21]. Therefore, treatment plans generally allow a margin sufficient for the organs to move. However, this causes a number of side effects, according to the respiratory displacement than the radiation dose is over delivered in normal tissue [20–23]. Radiation therapy has been studied by continuously considering regular breathing in order to solve this problem. As seen in the results of applying the rules of movement and irradiation studies for making the self-produced dynamic phantom by Jang et al., the side effects to the normal tissues can be reduced in order to increase the accuracy of the radiation treatment when improving the regularity of breathing [8].

The purpose of this study is to improve the regularity of respiration using the video coached respiration guiding system during RGRT. Using the real-time position management (RPM) system, the position is obtained in real time by the camera owing to the light from the infrared illumination diodes that is reflected at two points of the blocked infrared signals. This process is the implementation of free and guided breathing by using the video coached system. An assessment of the usefulness of the signals obtained using the video coached system is conducted [8, 20].

# II. MATERIALS AND METHODS

## 1. Subject of study and respiratory signal acquisition

The study proceeded with 16 lung cancer patients with tumors in the chest and the abdomen, with an infrared-maker block placed on the patients' abdomens at first to obtain a free breathing signal. Then, a guided breathing signal was obtained using in-house developed video coached systems for 3 s. The patients were able to maintain a comfortable, stable position during the guided breathing [20, 23]. The video coached respiration guiding system involved the movement of a red simulation bar for the regular inducement of inhalation and exhalation (Fig. 1). We fit the video coached systems for patients to see well (Fig. 2). In order to examine the degree of improvement in the regularity, we first obtained free-breathing data for 5 min and then obtained guided-breathing data for 5 min using the video coached system. Using the breathing signals for free and guided breathing, determined according to the period in each phase, divided by 10, along with the analysis mean and standard deviation, we recognized the correlation between the respiration and tumors [24–25].

**2. Real-time position management respiratory-gated system**

The patients' breathing and tumor movements were similar to the regularly improving accuracy and efficiency during the RGRT [8, 23–24]. In the results of a study with five lung cancer patients by Kang et al., the tumor was similar to breathing the period and movement when using the video coached system was confirmed that it has reduced displacement and the respiratory period of the respiratory variability [22]. Many studies have reported improvements in the accuracy and efficiency during RGRT using the video coached system [10, 12, 20, 23].

The RPM Respiratory Gating system (Varian Medical System, Alto, CA, USA) is generally used for analyzing movements in respiration. Using an infrared camera, this system observes the chest and the abdomen movements of the patient during inhalation and exhalation [26–28]. Around the infrared camera, infrared illumination diodes emit light, which is reflected at two points of the marker block and directed onto the patient's abdomen. RPM depends on the position of the marker block. Therefore, the marker blocks close to the isocenter of the patient must maintain their position during the radiation treatment. The marker block reflects infrared radiation onto the skin while the patient breathes, and the patient's breath is measured to determine the position of the marker block with respect to the reflected infrared camera in real time (Fig. 3.).

The RPM system can measure respiration by two modes: the amplitude mode, based on the location, and the phase mode, based on the period [8, 24–25, 29]. The amplitude mode is obtained if less or more breathing occurs than that in the reference point, which is not irradiated. The phase mode is obtained only at the phase set in the interval of 0 to 100%. It should also be a percentage (0–99%), so that it constantly cycles the period of the breath of the patient [20]. The phase mode, used in this study, increases the accuracy of the treatment by using the regular breathing of the signal after analyzing the movement of the tumor for the breathing and calculating the average and the standard deviation for the free breathing and guided breathing.

Irradiation is not accurately measured according to the change in the respiration using the RPM Respiratory Gating System in the phase mode, wherein the reference to the radiotherapy cannot be performed [8, 23–26]. We determine the influence on the accuracy of the RPM Respiratory Gating System according to the change in the respiration using the video coached system [27, 30].

**3. Analysis of the respiratory signal and change the phase by respiratory period**

We conducted an analysis of the free and guided breathing signals to improve the regularity of breathing using the video coached system by comparing the mean and standard deviation values according to the breathing cycles of each phase. First, we obtained the patients' respiratory signals using the phase mode of the RPM respiratory gating system. Here, there was a big change compared with the normal breathing during the acquisition of the respiratory signals, as indicated by the warning change from black to red on the display [27, 31]. If these warnings appeared during the guided breathing, this indicated a lack of concentration; i.e., the patient was breathing or as if asleep for one of several reasons when watching the video coached respiration guiding system. This change in the patient's breathing should be stabilized using an auditory system along with a visual system in order to maintain constant respiration.

This study poses a lower risk of irregular respiration because it does not involve actual radiation for obtaining a respiratory signal. The data obtained the file on the respiration signal and the structure of the data is shown in Figure 4. Data regarding the respiration movement are recorded on VXP files, which also contain the version of the system, the total recording time, etc. An important part of (1) is the maximum and minimum values of the respiratory signal; of (2), the values of the movement in the X-axis, Y-axis, Z-axis; of (3), to show the phase value in radians; and of (4), to record the total breathing times in milliseconds [27]. We determined the respiratory cycles using the data of the respiration signals, the radian of conversion to the degree, and divided them into phases of 10 steps: 0%, 10%, 20%, 30%, 40%, 50%, 60%, 70%, 80%, and 90%. We obtained the mean and standard deviation values for each phase using Microsoft Excel 2010.

## III. RESULTS

In this study, we evaluated the utility of the video coached respiration guiding system with 16 lung cancer patients. Using the data for 5 min of free and guided breathing, respiratory data was obtained for 5 min, and the mean and standard deviation of the respiration signal were calculated based on the phase. The mean and standard deviation values for each patient are shown in Table 1. The standard deviation exhibited a reduction of up to 59.85% during inhalation and a reduction of up to 71.08% during exhalation in the guided breathing compared with the free breathing. In the case of patient 6, the standard deviations for inhalation in the free breathing and guided breathing were 0.125cm and 0.064cm, respectively, and the values for exhalation were 0.231cm and 0.175cm, respectively (Table 1). This patient exhibited decreases of 51.36% and 75.78% in the standard deviation for exhalation in the free and guided breathing, respectively. However, the standard deviation values for patient 2 were increased by 31.11% and 36.03% during inhalation and exhalation, respectively, using the video coached system. It was found that the effect did not improve the regularity of the breathing compared with that of the other patients.

Figure 5 indicates the displacement of the breathing for the free breathing and the guided breathing using the video coached system. (A) represents the free breathing of patient 6, and (B) represents the same patient's guided breathing using the video coached respiration guiding system. Figure 6 shows the inhale waveform and average waveform for patient 6. Figure 7 shows the decrease in the standard deviation for the free and guided breathing. The period and standard deviation values for all of the patients' free and guided breathing are presented in Table 2. The average of the standard deviation decreased for the guided breathing, except for patient 11. The standard deviation was also reduced for the guided breathing, except for patient 8.

In this study, the period and standard deviation were reduced by guided breathing compared with those for free breathing for 12 patients. There was a 73% improvement in the regularity of the patients' breathing. The data for the patients' free and guided breathing were compared. The standard deviation increased the in case where the video coached respiration guiding system was used for a total of four patients. Depending on the patients were different in the process to analyzing the data of respiratory cycle and the respiration signal in accordance with different due to this. When using the video coached respiration guiding system of these results that different to improve the regularity of breathing for each patient [20, 23].

## IV. DISCUSSION AND CONCLUSION

The organ movement during breathing is very important in radiation therapy. Improving the regularity of breathing during radiation therapy can reduce the side effects on normal tissue [20, 27]. We used a video coached respiration guiding system in order to improve the efficiency and accuracy of the treatment, and we evaluated the system's usefulness. For 12 patients who used the video coached respiration guiding system, the standard deviation value was reduced, but for four patients, the standard deviation increased for guided breathing. In this case, the patient was expected to exhibit irregular breathing, coughing, swallowing saliva, etc. After the free breathing, we verified that the concentration of the patients who performed guided breathing using the video coached respiration guiding system was reduced over time, and the effect of the guided breathing was also reduced. In this case, we explained to the patient again how to breathe and the patient proceeded with the breathing, but sometimes complained of pain during the breathing, etc. The study poses several problems for actual lung cancer patients. Therefore, to enable more accurate study, a system is necessary to maintain the comfortable breathing of the patient for a long time.

Many treatments have been developed in the field of radiation therapy to supply as much radiation as possible to the target volume while minimizing the dose to the normal tissue. Therefore, it will be even more important in new radiation therapy and X-ray research to analyze respiratory movement using the video coached respiration guiding system reported in this study.

We demonstrated a technique to improve the regularity of breathing and determine the movement of normal tissue and tumors with 16 lung cancer patients using a video coached respiration guiding system. The system is expected to reduce error-treatment planning and treatment.

## ACKNOWLEDGEMENT

This work was supported by the Nuclear Safety Research Program through the Korea Radiation Safety Foundation (KORSAFe) and the Nuclear Safety and Security Commission (NSSC), Republic of Korea (Grant No. 1305033).

Tables

Table 1. Means and standard deviations of the inhale peak and exhale peak between free breathing and guided breathing for 16 patients. When the breathing was guided, the standard deviations at the end of exhale and inhale were reduced compared to the values for free breathing.

|  |  | P1 | | P2 | | P3 | | P4 | | P5 | | P6 | | P7 | | P8 | |
|---|---|---|---|---|---|---|---|---|---|---|---|---|---|---|---|---|---|
|  |  | Free | Guided | Free | Guided | Free | Guided | Free | Guided | Free | Guided | Free | Guided | Free | Guided | Free | Guided |
| Inhale Peak | Mean (cm) | 0.208 | -0.210 | 5.847 | 6.145 | 4.450 | 5.059 | 4.008 | 4.587 | 3.263 | 2.024 | 1.826 | 2.102 | -0.685 | 0.362 | 0.145 | 0.156 |
|  | ±SD (cm) | 0.180 | 0.101 | 0.075 | 0.169 | 0.391 | 0.163 | 0.071 | 0.028 | 0.054 | 0.053 | 0.125 | 0.064 | 0.098 | 0.100 | 0.139 | 0.094 |
| Exhale Peak | Mean (cm) | -0.297 | -1.045 | 5.563 | 5.307 | 3.974 | 4.276 | 3.349 | 3.458 | 2.265 | 0.756 | 1.129 | 0.976 | -1.388 | -0.836 | -1.322 | -1.048 |
|  | ±SD (cm) | 0.625 | 0.181 | 0.184 | 0.257 | 0.682 | 0.287 | 0.198 | 0.175 | 0.328 | 0.136 | 0.231 | 0.175 | 0.148 | 0.129 | 0.218 | 0.294 |

|  |  | P9 | | P10 | | P11 | | P12 | | P13 | | P14 | | P15 | | P16 | |
|---|---|---|---|---|---|---|---|---|---|---|---|---|---|---|---|---|---|
|  |  | Free | Guided | Free | Guided | Free | Guided | Free | Guided | Free | Guided | Free | Guided | Free | Guided | Free | Guided |
| Inhale Peak | Mean (cm) | 2.153 | 0.406 | -1.374 | -2.260 | 5.525 | 5.629 | -1.660 | -1.646 | 2.139 | 2.279 | 2.153 | 0.406 | -1.374 | -2.260 | 5.525 | 5.629 |
|  | ±SD (cm) | 0.175 | 0.061 | 0.109 | 0.062 | 0.116 | 0.049 | 0.089 | 0.068 | 0.084 | 0.063 | 0.175 | 0.061 | 0.109 | 0.062 | 0.116 | 0.049 |
| Exhale Peak | Mean (cm) | 1.400 | -0.653 | -2.595 | -3.789 | 5.008 | 2.913 | -2.302 | -2.494 | 1.865 | 1.813 | 1.400 | -0.653 | -2.595 | -3.789 | 5.008 | 2.913 |
|  | ±SD (cm) | 0.129 | 0.114 | 0.360 | 0.212 | 0.244 | 0.106 | 0.202 | 0.114 | 0.083 | 0.077 | 0.129 | 0.114 | 0.360 | 0.212 | 0.244 | 0.106 |

Table 2. Mean and standard deviation for one period (T) for each patient in free and guided breathing.

| Patient | Mean (s) | | ±SD (s) | |
|---|---|---|---|---|
| | Free | Guided | Free | Guided |
| 1 | 3.737 | 3.678 | 1.176 | 0.640 |
| 2 | 4.100 | 2.962. | 1.697 | 1.356 |
| 3 | 4.341 | 3.644 | 1.975 | 1.141 |
| 4 | 4.011 | 3.318 | 1.418 | 0.869 |
| 5 | 3.978 | 3.352 | 0.972 | 0.332 |
| 6 | 3.689 | 3.010 | 0.773 | 0.191 |
| 7 | 3.125 | 2.992 | 2.022 | 1.374 |
| 8 | 3.908 | 3.441 | 1.181 | 1.362 |
| 9 | 4.612 | 4.201 | 2.223 | 2.167 |
| 10 | 5.695 | 3.289 | 1.357 | 0.574 |
| 11 | 3.296 | 3.329 | 1.025 | 0.211 |
| 12 | 5.613 | 2.988 | 1.752 | 0.158 |
| 13 | 3.856 | 3.326 | 2.475 | 1.258 |
| 14 | 3.036 | 2.799 | 1.895 | 0.246 |
| 15 | 3.348 | 3.192 | 0.977 | 0.955 |
| 16 | 4.056 | 3.014 | 1.814 | 0.132 |

**List of figure captions**

Fig. 1. Video coached respiration guiding system.

Fig. 2. Patient in supine position.

Fig. 3. Real-time position management (RPM). (a) RPM respiratory gating system, (b) maker block, (c) RPM CCD camera surrounded by infrared light-emitting diodes (LEDs) and a display monitor.

Fig. 4. Structure of the recording files produced during the real-time position management (RPM) respiratory gating process (vxp file).

Fig. 5. Respiratory signal obtained from free breathing (A) and guided breathing (B).

Fig. 6. Each respiratory signal divided by inhale peak and average waveform. (A) Free breathing and (B) guided breathing.

Fig. 7. Average displacement and standard deviation determined at each phase of respiratory cycles. (A) Free breathing and (B) guided breathing.

**Figures**

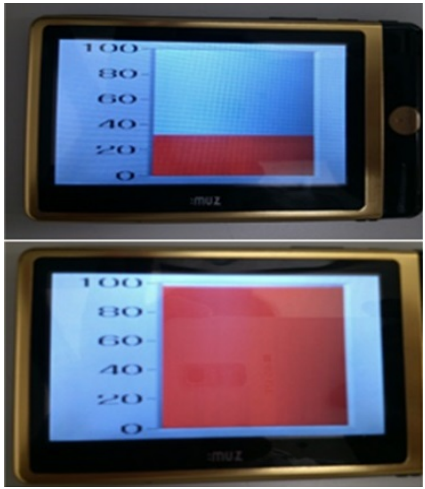

Fig. 1

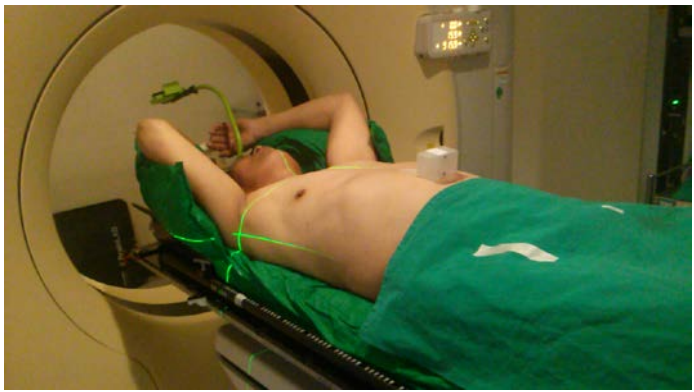

Fig. 2

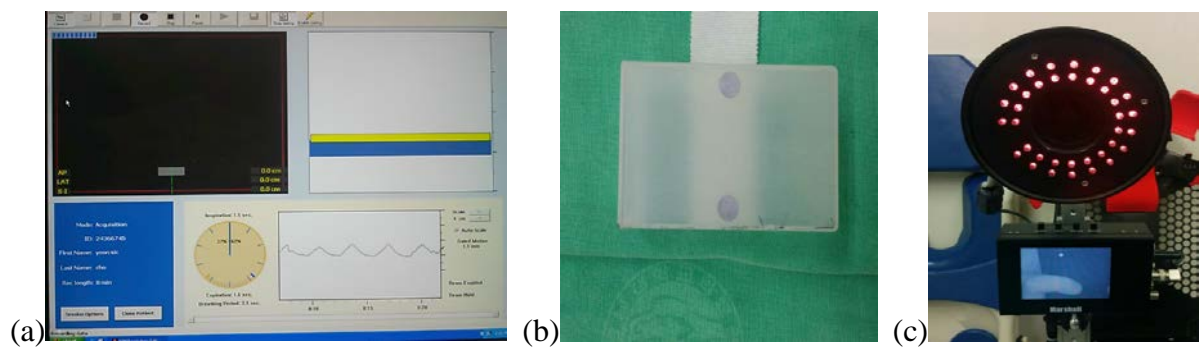

Fig. 3

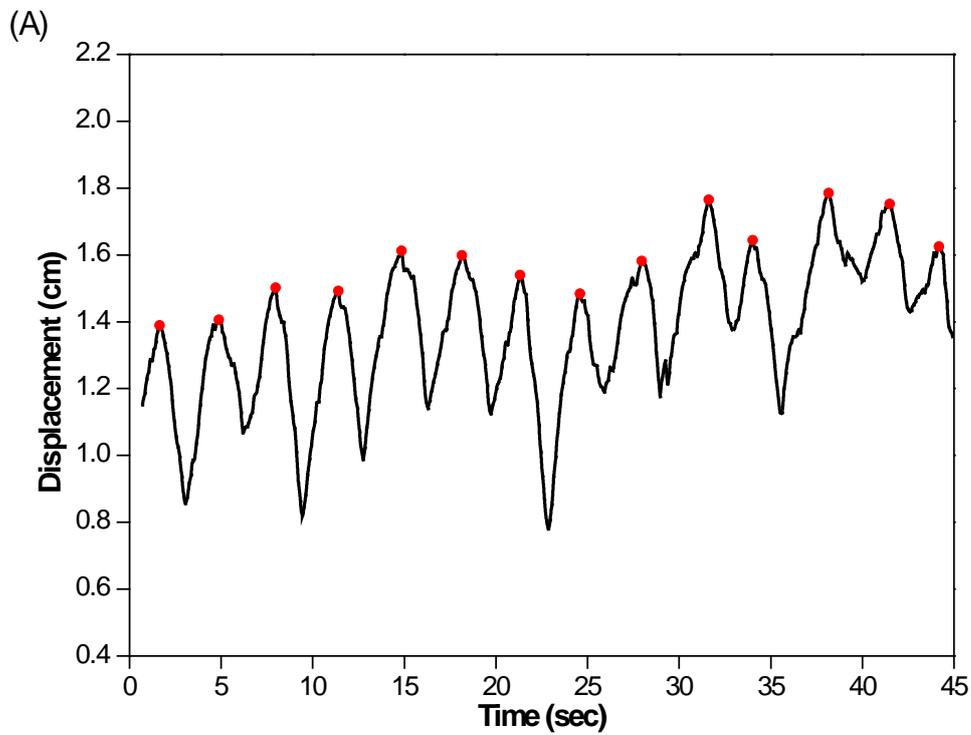

Fig. 4

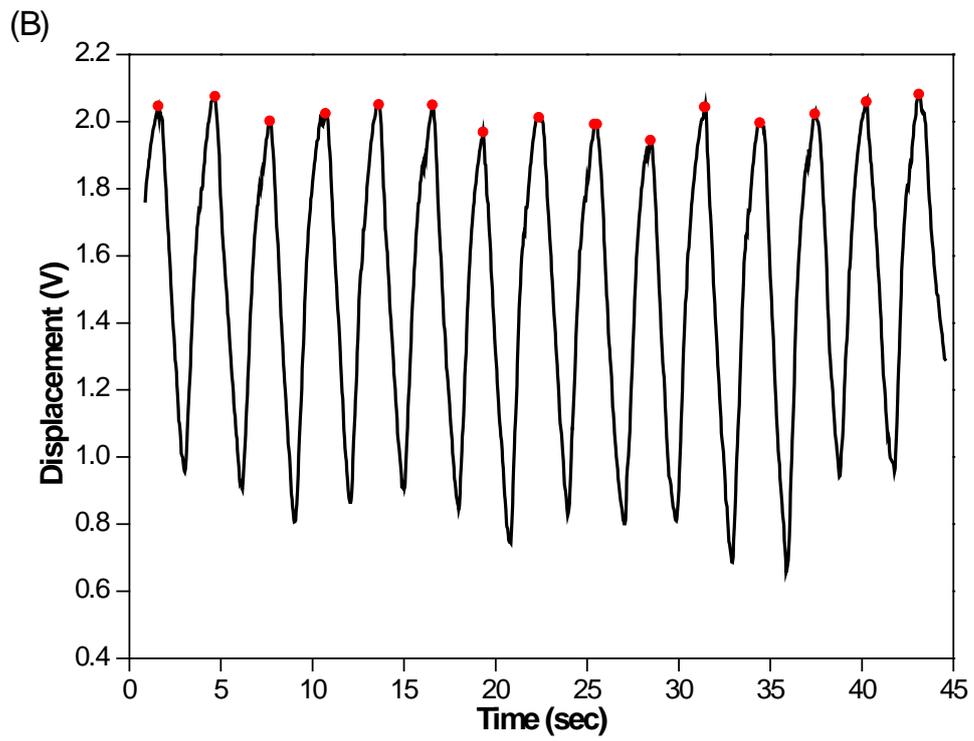

Fig. 5

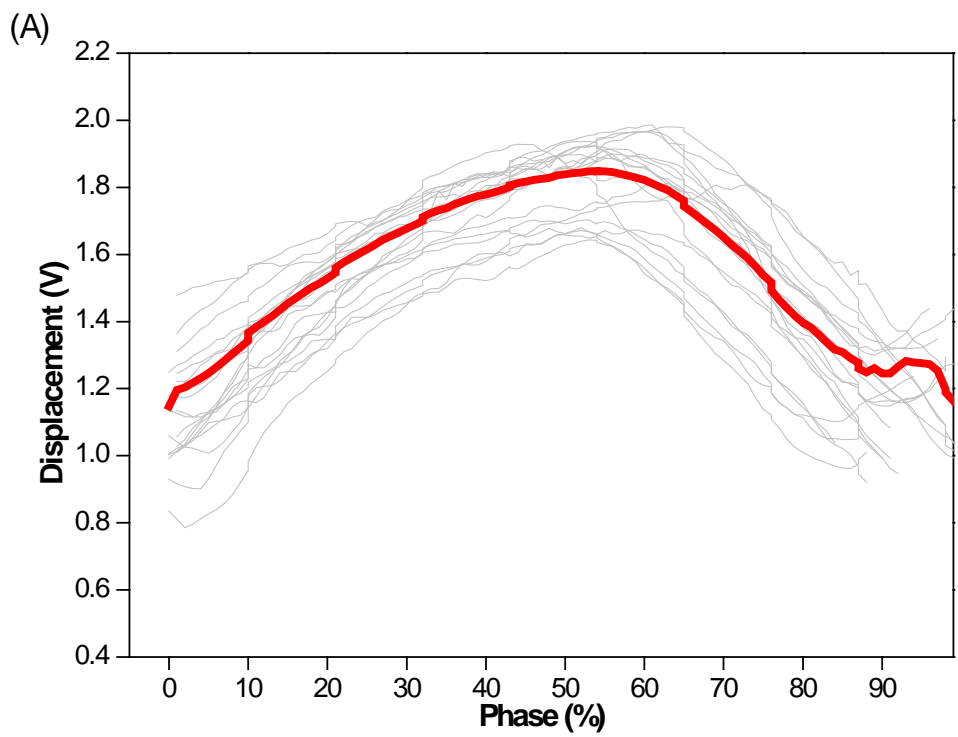

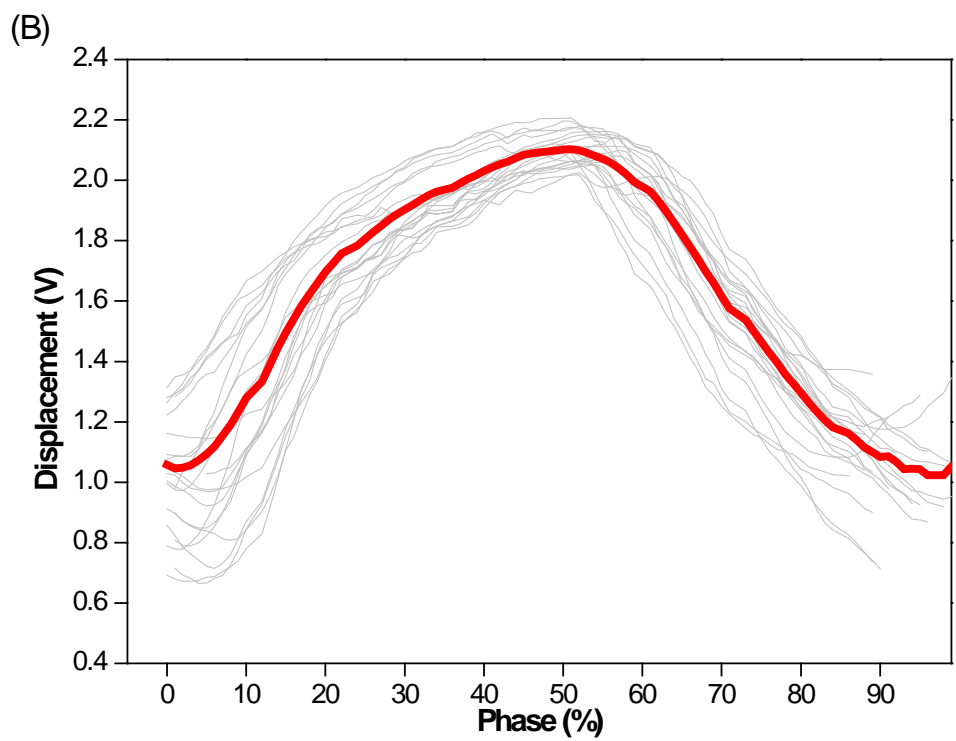

Fig. 6

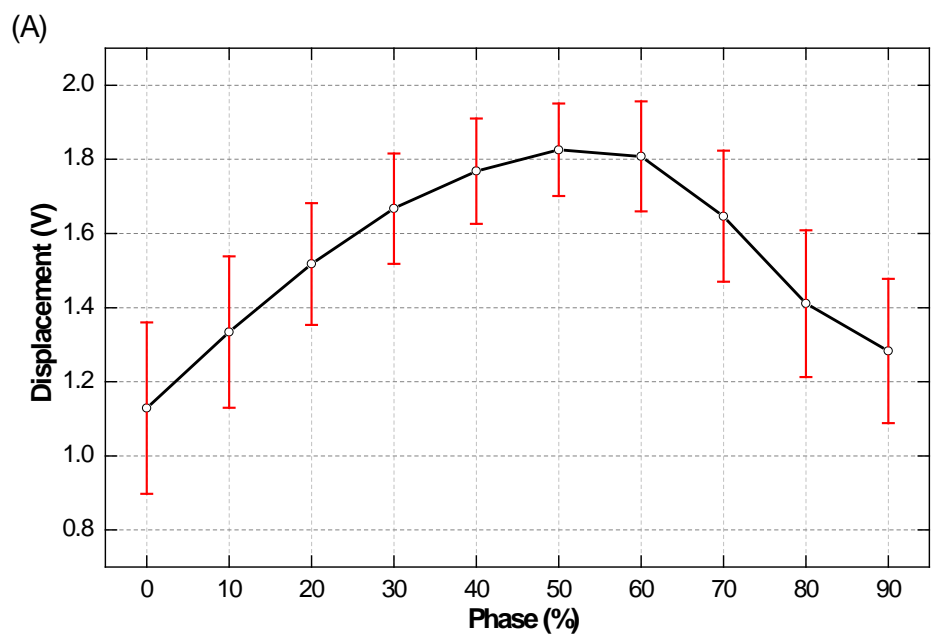

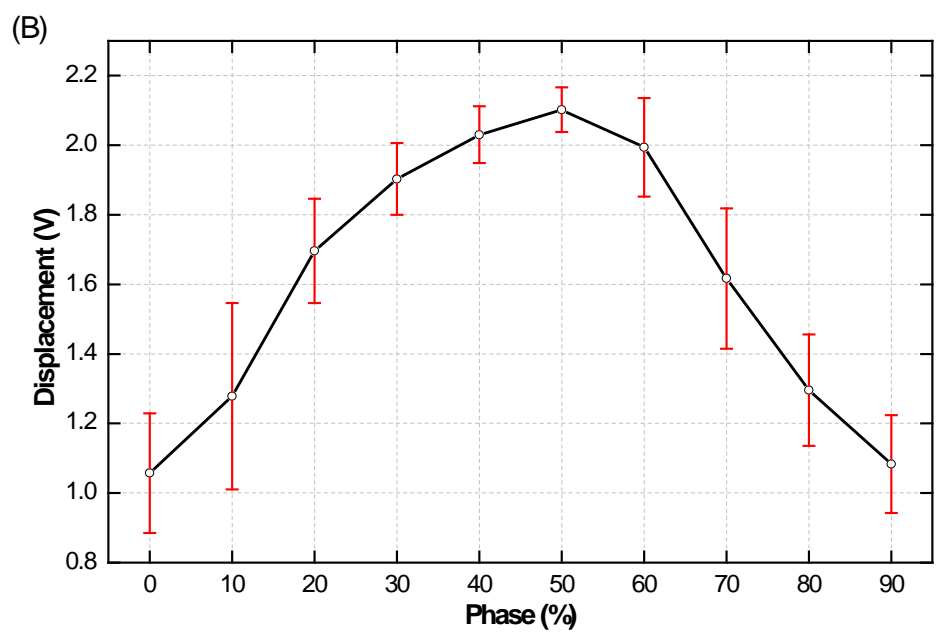

Fig.7